\documentclass[twocolumn,showpacs,preprintnumbers,amsmath,amssymb,floatfix]{revtex4}

\usepackage{graphicx}
\usepackage{dcolumn}
\usepackage{bm}

\begin{document}

%

\let\a=\alpha      \let\b=\beta       \let\c=\chi        \let\d=\delta
\let\e=\varepsilon \let\f=\varphi     \let\g=\gamma      \let\h=\eta
\let\k=\kappa      \let\l=\lambda     \let\m=\mu
\let\o=\omega      \let\r=\varrho     \let\s=\sigma
\let\t=\tau        \let\th=\vartheta  \let\y=\upsilon    \let\x=\xi
\let\z=\zeta       \let\io=\iota      \let\vp=\varpi     \let\ro=\rho
\let\ph=\phi       \let\ep=\epsilon   \let\te=\theta
\let\n=\nu
\let\D=\Delta   \let\F=\Phi    \let\G=\Gamma  \let\L=\Lambda
\let\O=\Omega   \let\P=\Pi     \let\Ps=\Psi   \let\Si=\Sigma
\let\Th=\Theta  \let\X=\Xi     \let\Y=\Upsilon

%

%

\def\cA{{\cal A}}                \def\cB{{\cal B}}
\def\cC{{\cal C}}                \def\cD{{\cal D}}
\def\cE{{\cal E}}                \def\cF{{\cal F}}
\def\cG{{\cal G}}                \def\cH{{\cal H}}
\def\cI{{\cal I}}                \def\cJ{{\cal J}}
\def\cK{{\cal K}}                \def\cL{{\cal L}}
\def\cM{{\cal M}}                \def\cN{{\cal N}}
\def\cO{{\cal O}}                \def\cP{{\cal P}}
\def\cQ{{\cal Q}}                \def\cR{{\cal R}}
\def\cS{{\cal S}}                \def\cT{{\cal T}}
\def\cU{{\cal U}}                \def\cV{{\cal V}}
\def\cW{{\cal W}}                \def\cX{{\cal X}}
\def\cY{{\cal Y}}                \def\cZ{{\cal Z}}

%

\newcommand{\Ns}{N\hspace{-4.7mm}\not\hspace{2.7mm}}
\newcommand{\qs}{q\hspace{-3.7mm}\not\hspace{3.4mm}}
\newcommand{\ps}{p\hspace{-3.3mm}\not\hspace{1.2mm}}
\newcommand{\ks}{k\hspace{-3.3mm}\not\hspace{1.2mm}}
\newcommand{\des}{\partial\hspace{-4.mm}\not\hspace{2.5mm}}
\newcommand{\desco}{D\hspace{-4mm}\not\hspace{2mm}}

\def\deltaakpi{\Delta {{A}}_{K\pi}}


\title{\boldmath 
$Y(4140)$: Possible options }

\author{Namit Mahajan
}
\email{nmahajan@prl.res.in}
\affiliation{
 Theoretical Physics Division, Physical Research Laboratory, Navrangpura, Ahmedabad
380 009, India
}


\begin{abstract}
We discuss possible options for interpreting the newly observed state $Y(4140)$
by the CDF collaboration in $B^+\to J/\psi\phi K^+$ decay above the $J/\psi\phi$ threshold, and argue
that it is more likely to be a $D_s^*$-$D_s^*$ molecular state or an exotic ($J^{PC}=1^{-+}$) hybrid charmonium.  
We have discussed decay modes which would allow unambiguous identification of the hybrid charmonium option. 

\end{abstract}

\pacs{
 }
\maketitle


A plethora of states have been observed which are not easy to accommodate within the
quark model picture of hadrons \cite{exotica}. Very recently, CDF collaboration has reported the evidence
of a narrow structure near the $J/\psi\phi$ threshold in $B^+\to J/\psi\phi K^+$ decay at $3.8\sigma$
\cite{Aaltonen:2009tz}. 
A fit to S-wave relativistic Breit-Wigner form for the narrow structure yields
a mass of $4143.0\pm 2.9\pm 1.2$ MeV and a width of $11.6^{+8.3}_{-5.0}\pm 3.7$ MeV. The new state, called $Y(4140)$ by the CDF collaboration,
has some similarity to the state $Y(3940)$ observed by the Belle collaboration \cite{Abe:2004zs}
 in $B\to K\,Y(3940)\to K J/\psi\omega $ modes (with a width of
 about $87$ MeV),  as far as production near $J/\psi\omega$
threshold is concerned. The same was confirmed by the BaBar collaboration \cite{Aubert:2007vj} with 
 the mass and width of
 $Y(3940)$ somewhat lower than those quoted by Belle \cite{Abe:2004zs}.
 This similarity could mean that $Y(4140)$ and $Y(3940)$ have similar
characteristics, and could even be partner states under a given scheme of quark assignments. In this note, we try to look for similarities and differences between the two states for each of the possible assignments discussed below, and identify the possible interpretations for $Y(4140)$ reported by the CDF collaboration.

Both $J/\psi$ and $\phi$ have $I^G(J^{PC})=0^-(1^{--})$ quantum numbers. Therefore, the newly
observed state, $Y(4140)$,  has positive C and G parities. Since the inferred 
mass is above the open charm threshold, this state will dominantly decay into open charm mesons with a 
large total width, if it is to be identified as a regular charmonium state. Therefore, it is highly unlikely to be one of the conventional
$c\bar{c}$ states. Other possibilities include a $[c\bar{c}s\bar{s}]$ tetraquark state, a molecular
structure or a hybrid charmonium state. Let us note in passing that a conventional charmonium assignment for 
$Y(3940)$, with it being identified as a $\chi_{c,J}^{(')}$ state, is still not ruled out \cite{Gonzalez:2009ma}.

In the tetraquark picture one expects the decays to proceed via the
rearrangement of constituent quarks, thereby implying a large width for the state $\sim {\mathcal{O}}(100)$ MeV. For a $[c\bar{c}s\bar{s}]$ state, decays to both the hidden and open charm final states are likely to be equally probable.
The observed width of $Y(4140)$ (${\mathcal{O}}(10)$ MeV) is too small (almost by an order of magnitude)
for it to be identified with the tetraquark picture assuming that the typical widths within tetraquark picture are ${\mathcal{O}}(100)$. This however may not be always true.
The authors in \cite{Drenska:2009cd} find a state with mass $3927$ MeV which decays to $J/\psi\omega$ and has quantum numbers $J^{PC}=0^{++}$. This can then be readily identified as $Y(3940)$. Also, from \cite{Drenska:2009cd} it is clear that widths smaller than MeV for tetraquark states are possible, in contrast to the naive expectation of large widths in the tetraquark scheme. It is therefore plausible that tetraquark interpretation is a likely possibility. 
However, within the same model, there is no state found around $4140$ MeV. We thus conclude for the present that $Y(4140)$ is not likely to be a tetraquark. However, a more detailed investigation of the tetraquark option should be undertaken to conclusively rule it out as a possibility for $Y(4140)$.

The mass of $Y(4140)$ is more than the $D_s^+$-$D_s^-$ threshold ($\sim 3937$ MeV) and also the 
$D_s^{+}$-$D_s^{*-}$ threshold ($\sim 4082$ MeV), but lower than 
the $D_s^{*+}$-$D_s^{*-}$ threshold ($\sim 4225$ MeV), making it a possible $D_s^{*+}$-$D_s^{*-}$ molecule. Let us recall that the
molecular states of charmed mesons were predicted by Tornqvist long ago \cite{Tornqvist:1991ks} (see also \cite{Voloshin:1976ap},\cite{De Rujula:1976qd},
\cite{Manohar:1992nd}, \cite{Wong:2003xk}). 
The binding energy
in such a case is $\Delta(D_s^{*+}D_s^{*-})\sim 80$ MeV. This value is very similar to $\Delta(D^{*}\bar{D}^{*})\sim 85$ MeV for a corresponding
molecular state identified with $Y(3940)$. Therefore, $Y(3940)$ and $Y(4140)$ may be $D_q^{*}\bar{D}_q^{*}$ ($q=u,d,s$) molecular states, and the similar decay
patterns $Y\to J/\psi\omega(\phi)$ definitely support this interpretation. A possible cause of worry for the molecular assignment in this case would be the rather large binding energy. However, it has long been known that for heavy mesons, binding energies $\sim {\mathcal{O}}(50)$ MeV can be expected \cite{Tornqvist:1991ks}. This is certainly true for $B^*$-$B^*$ systems and may not be very unlikely for the charmed mesons, particularly when the accompanying quark is the strange quark. These expectations are based on one pion exchange potential and one can expect some changes beyond this approximation. One would expect that the binding energies in case of charmed mesons with and without strange quark to be different because of large strange quark mass. However, it is not immediately clear whether this difference is going to be too large or not. For the present, we therefore assume that typical binding energies as above are plausible.  
For a $D_q^{*}\bar{D}_q^{*}$ molecular state, the simplest assignment will
be that of an S-wave ($L=0$) leading to $J=0,\, 2$ for $Y(4140)$ as the only possibilities, 
and the decays would proceed via rescattering. Again, similar to the tetraquark picture, decays to hidden and open
charm final states may be more or less equally probable. The observed width of $Y(4140)$ is significantly
smaller than the Belle result for $Y(3940)$ width \cite{Abe:2004zs}, though the two widths are roughly similar if BaBar value \cite{Aubert:2007vj} is used for the width of $Y(3940)$.
Using the product branching ratio (naively averaged over the Belle \cite{Abe:2004zs} and BaBar \cite{Aubert:2007vj} results) and
assuming that $BR(B\to Y(3940) K) < BR(B\to J/\psi K)$, one estimates the partial width 
$\Gamma(Y(3940)\to J/\psi\omega) > \mathrm{few}$ MeV (see Godfrey and Olsen in \cite{exotica}). If in future with better
statistics, the widths of $Y(4140)$ and $Y(3940)$ eventually turn out to be similar, then the molecular picture may be very compelling 
as the explanation for them. A significantly large difference between their widths will perhaps need some dynamic mechanism to explain the width difference, even when the decay patterns are very similar.
If the molecular picture is correct, then one should also expect other molecules like $D_q^{*}\bar{D}_{q'}^{*}$ to exist. For example, in the above picture,
a molecular state formed out of $D^{*+}$ and $D_s^{*-}$
would be expected to lie around $(m(D^{*+})+m(D_s^{*-})-80) \sim 4040$ MeV. Such a state could be searched
in exclusive B decays in association with a kaon, in exactly the same way as $Y(4140)$ and $Y(3940)$. The signal would be a resonance structure little above the $J/\psi K^*$ threshold. As eluded to above, a possible difficulty with the molecular assignment of various $D_q^{*}\bar{D}_{q'}^{*}$ states is the large binding energy. For $Y(4140)$, the binding energy, $E_B\sim 80$ MeV, corresponds to a size $< 0.4$ fm ($E_B\sim (2 M_{\mathrm{reduced}}R^2)^{-1}$), which is in the range where large strong interaction effects can not be simply ignored. Therefore, the wavefunction of $Y(4140)$ is unlikely to be dominated by the peripheral component, $\vert D_s^{*+}D_s^{*-}\rangle$,
but the multi-particle Fock states are expected to contribute significantly. If however, the molecular (peripheral) component dominates, one would expect an enhancement in $D_s^{*+}D_s^{*-}$ mass distribution near the threshold in $B \to D_s^{*+}D_s^{*-} K$
decays. This should be searched experimentally in order to better understand the underlying dynamics.
Apart from the hidden and open charm decays, within the molecular picture either of the mesons could decay radiatively yielding very characteristic decay patterns. More importantly, decay into two photons is possible with an appreciable width. \\

The possibility of $Y(4140)$ being a hybrid charmonium is very interesting. Within the flux-tube model, the low lying hybrid charmonium
states are expected to occur at $4.0 - 4.2$ GeV \cite{Isgur:1984bm}, \cite{Barnes:1995hc}. Lattice QCD predictions for the spin averaged
mass of the low lying hybrid charmonium states fall in the interval $4$-$4.4$ GeV \cite{latticehybrid}. 
$Y(4260)$ observed by the BaBar collaboration \cite{Aubert:2005rm} in the reaction 
$e^+e^-\to \gamma_{ISR}Y(4260)\to \gamma_{ISR}\pi^+\pi^-J/\psi$ has been strongly argued to be $1^{--}$
hybrid charmonium state \cite{y4260-hybrid}, as its mass ($\sim 4.26$ GeV) and width ($50$-$90$ MeV)
nicely fit in the hybrid picture while
other interpretations fall short of explaining the observed characteristics.
This would then imply existence of exotic hybrid charmonia as well around the same mass range. $Y(4140)$
could be the $1^{-+}$ exotic hybrid. Hybrid charmonium production in exclusive B decays has been studied \cite{Bdecayhybrid}. There is an interesting possibility that if $1^{-+}$ exotic hybrid lies
below the $D^{**}D$ threshold, then it can be a narrow resonance (see Lacock {\it etal} in \cite{latticehybrid} and Close and Page in \cite{Bdecayhybrid}). The decays of (at least exotic) hybrids
to a pair of S-wave ground state mesons are suppressed (see for example \cite{Page:1996rj}). Decays to one S-wave and another P- or D-wave meson are possible, which will give very distinctive signatures in the full angular distribution.
For a $1^{-+}$ exotic hybrid of mass about $4.1$ GeV, the estimated width is $10$-$15$ MeV (see Close and Page in \cite{Bdecayhybrid}) and one of the hidden charm final state is $J/\psi\phi$ (see Close {\it etal} in \cite{Bdecayhybrid}). All these estimates strongly suggest that $Y(4140)$ reported by the CDF collaboration is indeed consistent with  $J^{PC}=1^{-+}$ exotic hybrid charmonium. Following a similar line of reasoning, one may attempt to interpret $Y(3940)$ also as a hybrid charmonium. However, it seems to lie too low compared to the flux model and lattice QCD predictions. Also, the width of $Y(3940)$ seems to be relatively large to be easily identified with an exotic hybrid charmonium state. The hybrid interpretation of $Y(3940)$ has also been questioned by Close and Page \cite{y4260-hybrid}. Therefore, it is possible that though there is some similarity in the observed decay patterns of $Y(3940)$ and $Y(4140)$, they are very distinct in nature. Going by the above hints, $Y(4140)$ could be a genuine $1^{-+}$ exotic hybrid.
If it is so, then bulk of the width will be due to decay into light hadrons, and $DD^*$ final state will be an important open charm mode to search for (see Close and Page in \cite{Bdecayhybrid}). In the hidden charm category, decays to $\eta_c\eta'$ and radiative decay to $J/\psi$ seem promising.  {\em It is worthwhile to mention that the decay to two photons is forbidden}. For the 
$1^{-+}$ assignment, this can be seen as a consequence of Yang's theorem. Decay of a hybrid to two photons is generically
forbidden, see for example \cite{Page:1996qh}. This feature could be used to distinguish the hybrid assignment from any other possible assignments for $Y(4140)$, in particular the molecular picture. Another distinguishing feature is the decay to $J/\psi\omega$ final state, which is expected to be highly suppressed in the molecular picture. It is important to search for this decay mode to further establish the hybrid nature of $Y(4140)$.
Furthermore, the hybrid charmonium state could decay to non-strange charmed mesons, modes which are either forbidden or suppressed in the molecular interpretation. As has been remarked above, $Y(4260)$ has been strongly advocated to be a $1^{--}$ hybrid charmonium state. Therefore radiative transition $Y(4260) \to Y(4140) \gamma$ with a width of upto several tens of KeVs is possible, see \cite{Close:2002ky}. The reported  signal for $Y(4140)$ is the narrow peak in $J/\psi\phi$ invariant mass in $B^+ \to Y(4140) K^+$ mode. If $Y(4140)$ is indeed a hybrid charmonium (with $c\bar{c}g$ as constituents), it would also be produced at the hadron colliders in association with $J/\psi$ via $gg\to Y(4140)J/\psi$. This should be possible at Tevatron and more so at LHCb. If the hybrid interpretation turns out to be true, this would open up a totally new regime of quarkonium spectroscopy. It would also be interesting to search for similar bottom hybrids and hybrid baryons.

In this short note we have examined possible options for interpreting the newly reported narrow state, $Y(4140)$ just above the $J/\psi\phi$ threshold \cite{Aaltonen:2009tz}. It is not likely to be a conventional charmonium state or a tetraquark state. The molecular interpretation with it being a $D_s^{*+}$-$D_s^{*-}$ is quite tempting, particularly since there is a possibility to identify $Y(3940)$ as a $D^{*}$-$\bar{D}^{*}$ molecule. Many other partner states $D_q^{*}\bar{D}_{q'}^{*}$ ($q,q'=u,d,s$) are expected to lie in the intermediate mass range and should show up as resonances in $J/\psi\rho(K^*)$ modes. 
The other possibility explored is that of $Y(4140)$ being the $1^{-+}$ exotic hybrid. Various estimates within the flux tube models and lattice QCD calculations seem to strongly suggest that it is $1^{-+}$ hybrid charmonium. Being below the $D^{**}D$ threshold may make it narrow enough, which seems to be in line with the preliminary width determinations. 
More importantly, if $Y(4140)$ is $1^{-+}$ exotic hybrid then the following modes will play crucial role in clearly identifying the state and distinguishing from the molecular assignment: (i) it can decay to $J/\psi\omega$ final state and non-observation of it will seriously question the hybrid interpretation, (ii) two photon final state is not possible for a hybrid state while decay to two photons with an appreciable width is expected for the molecular picture, (iii) decays to non-strange charmed mesons (and also light hadrons) are expected to be suppressed in the molecular picture while such final states are easily reached if $Y(4140)$ is a hybrid, (iv) angular distribution of the decay products will show different and distinct threshold behaviour for the hybrid and molecular assignment. We also find it quite plausible that despite apparent similarity in the observed decay patterns of $Y(3940)$ and $Y(4140)$, these two states are intrinsically of different character.
To establish the quantum numbers and other properties of $Y(4140)$, it could possibly also be searched in the reaction $e^+e^-\to Y(4140)+ \gamma + n(\pi)$ and decay angular correlations studied. 

\vskip 0.25cm

While this note was being finalised, \cite{Liu:2009ei} appeared where the authors suggest that the molecular picture, similar to what is suggested here, is the preferred interpretation. They however feel that the hybrid charmonium interpretation is not the correct one. We tend to differ with this opinion. Till various properties and decay characteristics of both $Y(3940)$ and $Y(4140)$ are better understood, it may be premature to rule out the hybrid scenario for $Y(4140)$.



%

\end{document}